\documentstyle[12pt,blois,epsf]{article}

\newcommand{\msun}{\,M_{\sun}\,}

\def\msun{\ifmmode{\ {\rm M}_\odot}\else{$ {\rm M}_\odot$}\fi}  
\def\msunyr{\ifmmode{\msun \ {\rm yr}^{-1}}\else{$\msun \ {\rm 
yr}^{-1}$}\fi}

\begin{document}

\heading{HIGH RESOLUTION X-RAY SPECTROSCOPY OF GALAXIES AND 
          CLUSTERS WITH AXAF}

\author{Michael W. Wise} {Massachusetts Institute for Technology, 
       Center for Space Research, \\ AXAF Science Center, 
       Cambridge, MA, USA \\ wise@space.mit.edu }  {~}

\begin{bloisabstract}
We review the high resolution X--ray spectroscopy capabilities of the
AXAF observatory focusing on the High Energy Transmission Grating (HETG).
As part of the Guaranteed Time Observation (GTO) program, the HETG
science team will observe both elliptical galaxies and clusters of
galaxies. We discuss the problems associated with observing extended
sources with the HETG and some of the potential scientific insights
which AXAF's spectroscopy capabilities can provide for these classes
of objects. 
\end{bloisabstract}

\section{Introduction}
\label{sec:intro}

The High Energy Transmission Grating (HETG) onboard AXAF will provide
the first opportunity since the Einstein Focal Plane Crystal
Spectrometer (FPCS) to obtain high resolution X-ray spectra from the
cores of galaxies and clusters.  
These objects are known to contain large quantities of cool gas and
are rich sources of X-ray emission. 
With an energy resolution of  $E/\Delta E \sim 1000$, the HETG will be
capable of resolving individual X-ray emission lines, absorption
lines, and absorption edges in galaxy and cluster spectra. 
Using plasma diagnostics, these spectra will provide information on
source properties such as temperatures, ionization states, densities,
velocities, elemental abundances, and thereby structure, dynamics, and
evolution.
In this paper, we briefly review the technical capabilities
of the HETG and ACIS spectrometers onboard AXAF. 
With these capabilities in mind, we then discuss some of the
scientific issues which HETG observations of galaxies and clusters
will allow us to explore.

\section{Overview of the HETG}
\label{sec:hetg}

\begin{figure}[t]
\hbox{
  \epsfxsize=3.25in \epsfbox{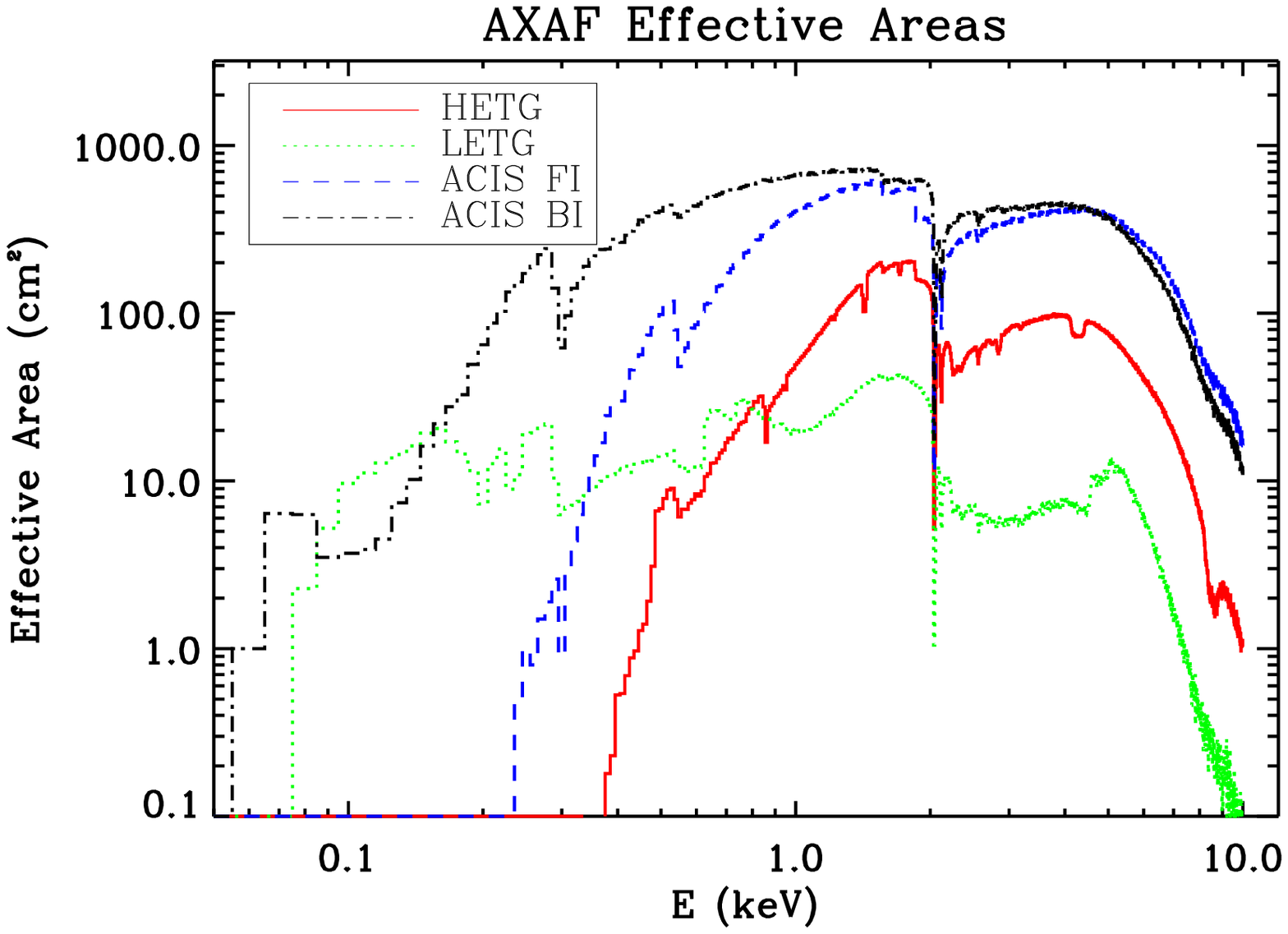}
  \hspace{0.1cm}
  \epsfxsize=3.25in \epsfbox{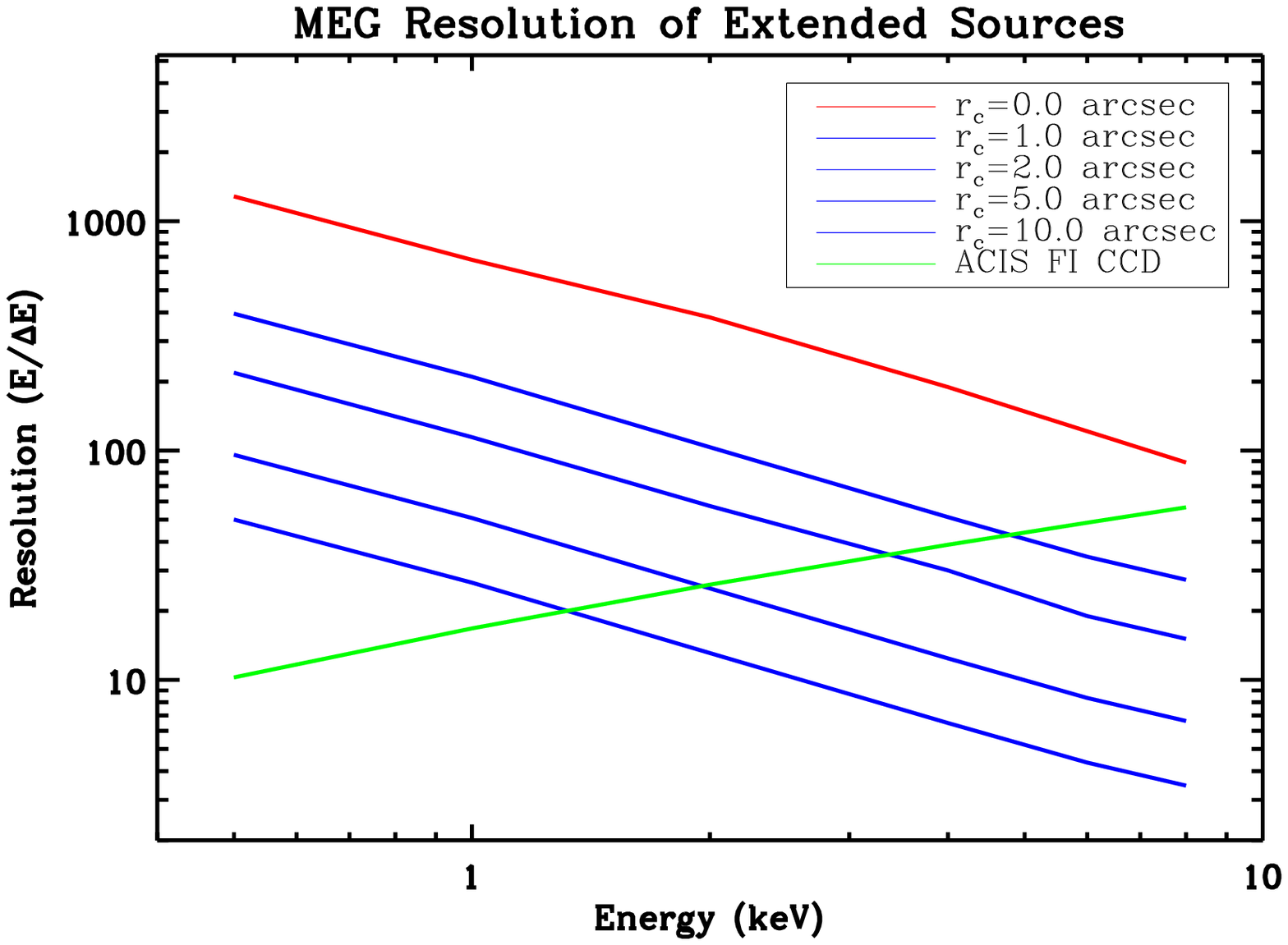}
     }
\caption[]{ \small a) The effective area for the primary science instruments
		onboard AXAF. Curves are shown for the FI and BI CCDs
		as well as the HETGS and LETGS grating first orders.
		b) The HETG spectral resolution as a function of the
		source's spatial extent. The source was modeled as a
		standard cluster beta model with a core radius $r_c$.
		The curve for a point source is shown for comparison
		as is the intrinsic resolution of the ACIS CCD.
          }
\label{fig:ea}
\end{figure}

The HETG is one of four scientific instruments (SIs) that will operate
onboard NASA's Advanced X-ray Astrophysics Facility (AXAF).
These instruments include two imaging detectors: the AXAF CCD Imaging
Spectrometer (ACIS) and the High Resolution Camera (HRC).  
The remaining two instruments consist of dispersive grating
spectrometers: the HETG and LETG (Low Energy Transmission Grating).
The preferred detector for the HETG is the ACIS-S 
consisting of 6 imaging CCDs in a 1x6 array.  
Figure~\ref{fig:ea}a) presents the effective areas for the primary
science instruments onboard AXAF. 
By combining the AXAF mirror's high angular resolution ($\sim 1$ arcsec) 
and the grating's large diffraction angles ($\sim 100$ arcsec/\AA), 
the HETG provides spectral resolutions up to $E/\Delta E \sim
1000$ in the AXAF energy band.
The HETG consists of 336 individual grating facets of two types: High
Energy Grating (HEG) facets and Medium Energy Grating (MEG) optimized
for maximal efficiency above and below 2.0 keV, respectively.  
Table~\ref{tab:hetg} summarizes the spectral capabilities of the 
HETG as well as the native spectroscopic abilities of the ACIS
detector itself.
Details of the HETG have been presented previously in SPIE proceedings
and various AXAF Science Center (ASC) publications~[1,2].

%
%

\begin{table}[b]
\small
\caption{X-ray Spectroscopy with AXAF}
\begin{center}
\begin{tabular}{|lllll|}
\hline 
\multicolumn{5}{|l|}{High Energy Transmission Grating ({\bf HETG})} \\
 &                & MEG                 & HEG               & \\
 & Grating Period & 4001.41 ang         & 2000.81 ang   & \\
 & Energy Range   & $0.4-5.0$ keV       & $0.9-10.0$ keV    & \\
 & Resolution     & $E/\Delta E = 520$ \@ 1.0 keV  & (point sources) &  \\
 &                & $E/\Delta E = 1000$ \@ 1.0 keV & (point sources) &  \\
\multicolumn{5}{|c|}{~} \\
\multicolumn{5}{|l|}{AXAF CCD Imaging Spectrometer ({\bf ACIS})} \\
 & Imaging array  & 4 CCDs (all FI) & FOV=$16.9\times 16.9$ arcmin & \\
 & Spectroscopy array  & 6 CCDs (4 FI, 2 BI) & FOV=$8.3\times 50.6$ arcmin & \\
 & Resolution     & FI: $E/\Delta E = 10, 46$ \@ 0.5, 5.9 keV  & &  \\
 &                & BI: $E/\Delta E = 4.3, 31$ \@ 0.5, 5.9 keV  & &  \\
\hline 
\end{tabular}
\end{center}
\label{tab:hetg}
\end{table}

\begin{figure}[t]
\hbox{
  \epsfxsize=3.25in \epsfbox{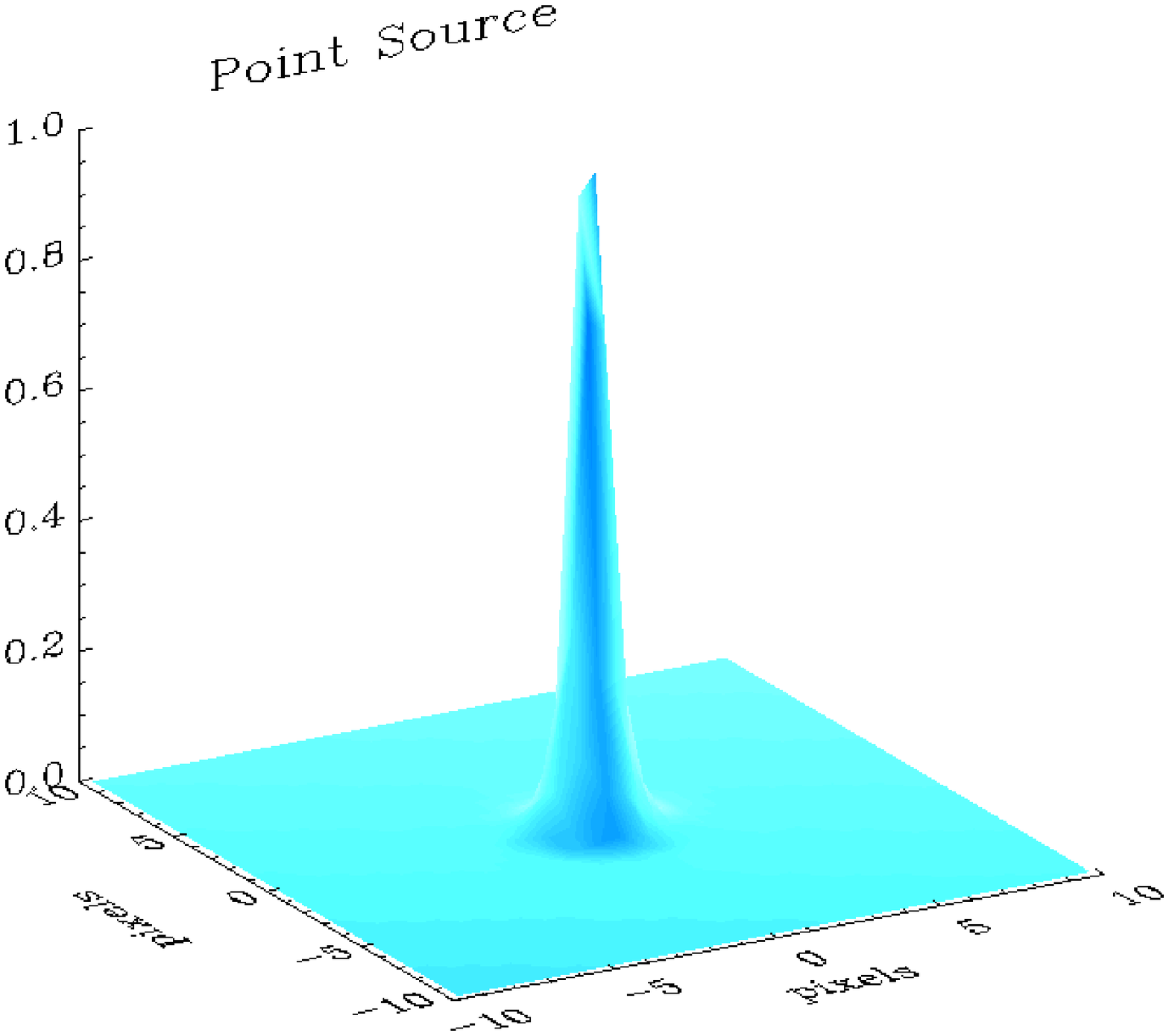}
  \hspace{0.1cm}
  \epsfxsize=3.25in \epsfbox{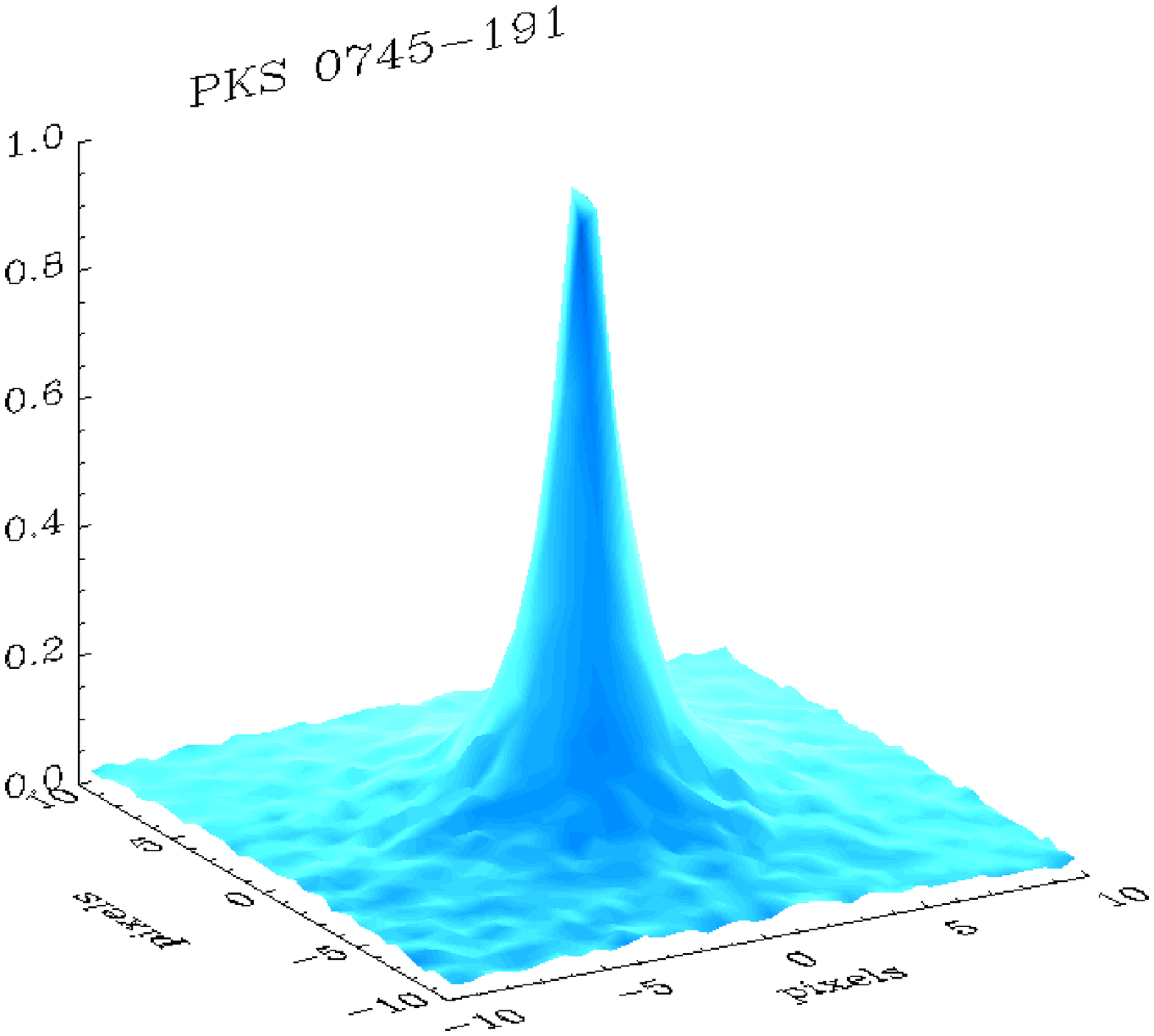}
     }
\caption[]{ \small A comparison of the surface brightness
	    distribution for the cooling flow PKS0745-191 and the
	    intrinsic width of the AXAF point response function.
	    The AXAF PSF is shown on the left and PKS0745-191 is
	    shown on the right. The X and Y axes are expressed in
	    units of ACIS pixels. 
          }
\label{fig:psf}
\end{figure}

The HETG was designed to provide optimal spectral resolution for 
point sources. If the observed object has a significant spatial
extent, the spectral resolving power ($E/\Delta E$) of the HETG
will be degraded. Figure~\ref{fig:ea}b) depicts the effects of
increasing source extent on the HETG spectral resolution.
For objects with core radii less than $\sim 5$ arcsec, the spectral 
resolving power of the HETG can still exceed that of the ACIS imager
for energies below about 2.0 keV.  
In the case of cooling flow clusters and ellipticals at Virgo
distances, the X--ray emission is expected to be quite
peaked (sharp surface brightness distributions are one of the 
{\it signatures} of cooling flows), producing even 
higher spectral resolution. 
This point is illustrated in Figure~\ref{fig:psf} which compares
the intrinsic AXAF point response function (PSF) with the surface
brightness distribution of the cluster PKS0745-191.
Thus, the Fe L emission from the ISM in elliptical galaxies and the
ICM in clusters of galaxies can still be profitably explored with 
the HETG.

%
%

\section{Clusters of Galaxies}
\label{sec:clust}

X--ray observations of galaxy clusters indicate that the
material in the cores consists of a multiphase plasma. 
In many clusters, large amounts of intracluster gas appear to be
cooling below X--ray emitting temperatures with typical cooling
rates of 100--1000 M$_{\odot}$ yr$^{-1}$~[3].
Evidence for cooling in cluster cores include sharply peaked surface
brightness profiles, cooling times less than the estimated age of the
cluster, and the observation of low temperature, low ionization X--ray 
lines.
This last point is of particular importance, since it
represents direct evidence that cooler ($T \sim 10^{6-7}$ K) gas
is present at the centers of clusters~[4]. 
However, the fate of the cooling material in cluster cores is unknown,
and indeed that the gas is actually cooling and remaining below X--ray 
temperatures has not been proven. 
The ultimate repository of this cooling gas remains one of the most
important observational problems concerning galaxy clusters.

High resolution X--ray spectra of clusters with the HETG will allow us
to directly probe the physical state of material in the core. 
Measurement of individual line fluxes will provide several {\it 
independent} and {\it simultaneous} determinations of the 
amount of cooling material in these objects.
The detection of lines produced by material at several 
different densities and temperatures would verify that the 
gas in cluster cores is inhomogeneous as implied by 
broad-band X-ray surface brightness profiles. 
X-ray line profiles can provide information on opacity in the core and 
line widths will allow us to place limits on the turbulence 
in the gas which is likely to dominate over broadening due 
to inflow velocity or thermal broadening. Finally, 
accumulated cold material in cluster cores should produce a 
number of effects (X-ray line flux reduction, absorption 
edges, etc.) which are potentially observable with AXAF's 
high energy resolution~[5]. 
As part of the Guaranteed Time Observation (GTO) program, 
the HETG science is currently planning to observe two cooling
flow clusters: PKS0745-191 and Abell 1835.

\section{Elliptical Galaxies}
\label{sec:egal}

\begin{figure}[hb]
\epsfxsize=6.5in
\epsfysize=2.9in
\epsfbox{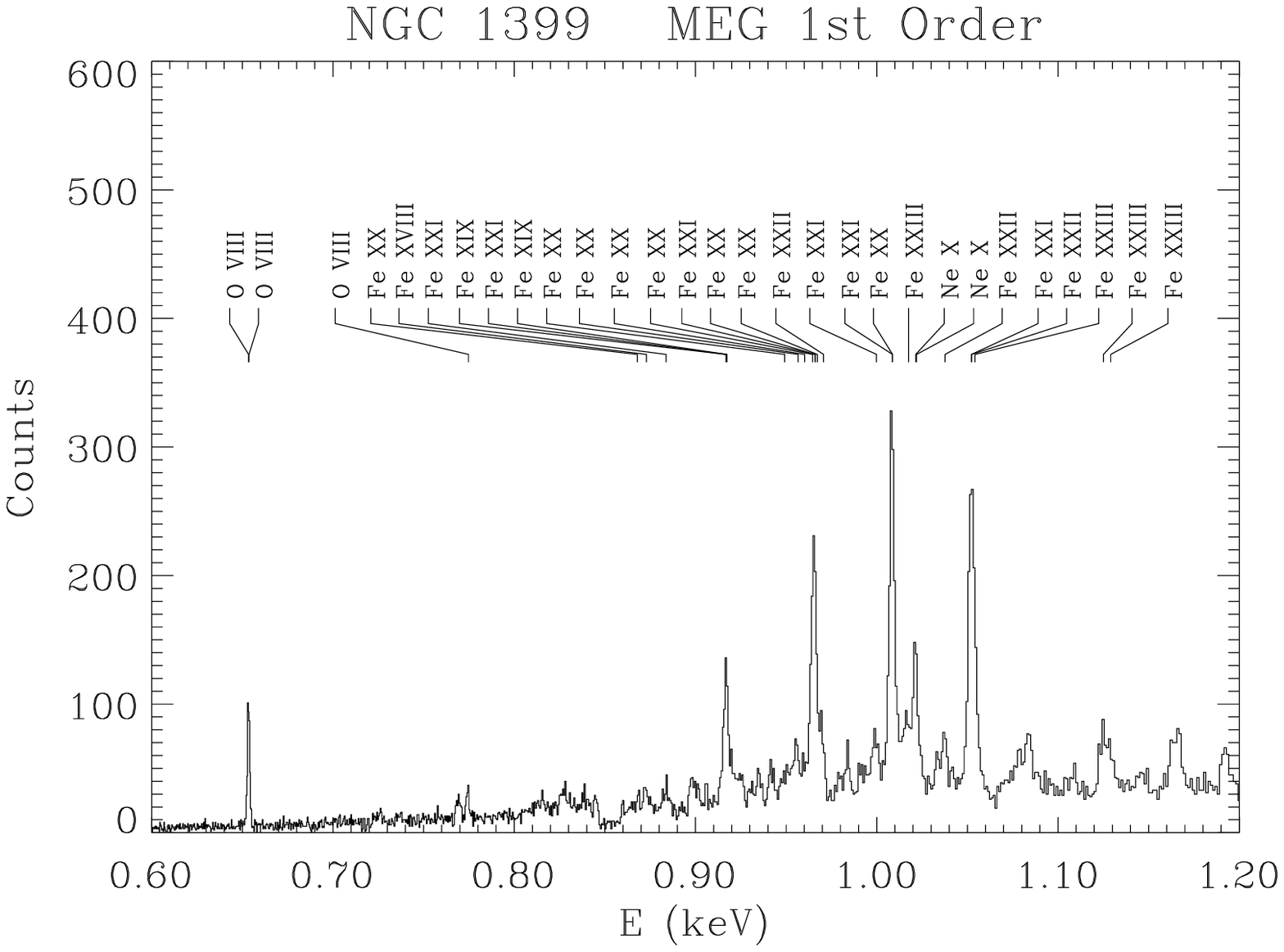}
\caption[]{\small Simulated HETG 1st order MEG spectrum for a 60 ksec
	   observation of NGC1399. The input spectral model was a
	   1.0 keV Raymond--Smith thermal plasma with 0.2 solar
	   abundances. Some of the stronger X--ray lines are indicated.}
\label{fig:egal}
\end{figure}

Elliptical galaxies are bright sources of X-ray emission
and contain large amounts of hot, interstellar gas.
For brighter X-ray galaxies, the inferred masses of hot gas are
consistent with those expected given the present rates of stellar 
mass loss. 
Spectral observations of elliptical galaxies indicate that the X-rays
are produced by thermal emission from diffuse gas at temperatures of
$T \approx 10^7$ K ($kT \approx 1$ keV).
Thus, like clusters, these objects should be rich sources of X--ray
line emission. 
Figure~\ref{fig:egal} shows the Fe L region for a simulated 60 ksec
HETG observation of the elliptical galaxy NGC1399. 
Bright emission lines from various elements are clearly resolved.
Current models of the X--ray emission from ellipticals require a low
rate of Type Ia supernova heating and chemical enrichment in the gas.
HETG observations such as these will place strong constraints on the
gas abundances in these objects.
In addition, for the brightest X-ray galaxies, the cooling times in
the gas are short, which suggests that the gas may form galactic
cooling flows. If present, strong X--ray line emission from this
cooling material should also be detectable with the HETG in many
ellipticals.

%
%

\vspace{-0.25in}
\acknowledgements{The HETG Science Team consists of 
C. R. Canizares (PI), C. Baluta, D. Davis,                    
J. E. Davis,                  
D. Dewey,                      
T. Fang,                    
K. A. Flanagan,            
A. Hicks,                   
J. Houck,                     
D. Huenemoerder,             
J. Kastner,                   
H. L. Marshall,             
N. Schulz,                 
M. D. Stage, and               
M. W. Wise                
}

%
%

\vfill
\end{document}